\pdfoutput=1

\documentclass[11pt,twoside,a4paper,cmspaper,final,collab]{cms-tdr}
\def\svnVersion{395955}\def\cmsCernNoTag{CERN-EP/2016-236}\def\cmsCernDate{\today}\def\cmsMessage{Published in Physical Review Letters as \href{http://dx.doi.org/10.1103/PhysRevLett.118.122301}{\doi{10.1103/PhysRevLett.118.122301.}}}

\begin{document}\cmsNoteHeader{HIN-16-009}

\hyphenation{had-ron-i-za-tion}
\hyphenation{cal-or-i-me-ter}
\hyphenation{de-vices}
\RCS$Revision: 385069 $
\RCS$HeadURL: svn+ssh://svn.cern.ch/reps/tdr2/papers/HIN-16-009/trunk/HIN-16-009.tex $
\RCS$Id: HIN-16-009.tex 385069 2017-02-04 20:58:13Z alverson $
\newlength\cmsFigWidth
\ifthenelse{\boolean{cms@external}}{\setlength\cmsFigWidth{0.98\columnwidth}}{\setlength\cmsFigWidth{0.4\textwidth}}
\ifthenelse{\boolean{cms@external}}{\providecommand{\cmsLeft}{top\xspace}}{\providecommand{\cmsLeft}{left\xspace}}
\ifthenelse{\boolean{cms@external}}{\providecommand{\cmsRight}{bottom\xspace}}{\providecommand{\cmsRight}{right\xspace}}
\providecommand {\roots}    {\ensuremath{\sqrt{s}}\xspace}
\providecommand {\rootsNN}  {\ensuremath{\sqrt{\smash[b]{s_{_\mathrm{NN}}}}}\xspace}
\newcommand {\deta}     {\ensuremath{\abs{\Delta\eta}}\xspace}
\newcommand {\pp}    {\ensuremath{\Pp\Pp}\xspace}
\newcommand {\PbPb}  {\ensuremath{\mathrm{PbPb}}\xspace}
\newcommand {\AuAu}  {\ensuremath{\mathrm{AuAu}}\xspace}
\newcommand {\pPb}  {\ensuremath{\Pp\mathrm{Pb}}\xspace}
\newcommand {\AonA}  {\ensuremath{\mathrm{AA}}\xspace}
\newcommand{\noff}    {\ensuremath{N_\text{trk}^\text{offline}}\xspace}
\cmsNoteHeader{HIN-16-009}
\title{Observation of charge-dependent azimuthal correlations in \texorpdfstring{\pPb}{pPb} collisions and its implication for the search for the chiral magnetic effect}

\date{\today}

\abstract{
Charge-dependent azimuthal particle correlations with respect to the second-order
event plane in \pPb and \PbPb  collisions at a nucleon-nucleon center-of-mass energy of 5.02\TeV have been studied
with the CMS experiment at the LHC. The measurement is performed with a three-particle
correlation technique, using two particles with the same or opposite charge within the pseudorapidity range
$\abs{\eta}<2.4$, and a third particle measured in the hadron forward calorimeters ($4.4<\abs{\eta}<5$). The observed differences
between the same and opposite sign correlations, as functions of multiplicity and $\eta$
gap between the two charged particles, are of similar magnitude in \pPb and
\PbPb collisions at the same multiplicities. These results pose a challenge for the
interpretation of charge-dependent azimuthal correlations in heavy
ion collisions in terms of the chiral magnetic effect.
}

\hypersetup{%
pdfauthor={CMS Collaboration},%
pdftitle={Observation of charge-dependent azimuthal correlations in pPb collisions and its implication for the search for the chiral magnetic effect},%
pdfsubject={CMS},%
pdfkeywords={CMS, physics, chiral magnetic effect, CME, small systems}}

\maketitle

In relativistic heavy ion collisions, metastable domains of gluon fields may form with non-trivial topological configurations~\cite{Lee:1973iz,Lee:1974ma,Morley1985,Kharzeev:1998kz}. The interaction of quarks with these gluon fields will lead to an imbalance in left- and right-handed quarks, which violates local parity ($P$) symmetry~\cite{Morley1985,Kharzeev:1998kz}.
In the presence of a strong magnetic field in a noncentral nucleus-nucleus (\AonA) collision, this chirality imbalance leads to an electric current perpendicular to the reaction plane, resulting in a
final-state charge separation phenomenon,
known as the chiral magnetic effect (CME)~\cite{Kharzeev:2004ey}.
Attempts to measure this charge separation in heavy ion collisions
were made by the STAR experiment at
RHIC~\cite{Abelev:2009ac,Abelev:2009ad,Adamczyk:2013kcb,Adamczyk:2014mzf,Adamczyk:2013hsi}
and the ALICE experiment at the LHC~\cite{Abelev:2012pa}.
In these measurements, a charge dependence of azimuthal correlations
with respect to the reaction plane was observed, which is qualitatively
consistent with the expectation of a charge separation from the CME.

The charge separation can be characterized by the $P$-odd sine term ($a_{1}$)
in a Fourier decomposition of the particle azimuthal distribution~\cite{Voloshin:2004vk}:
\begin{linenomath}
\ifthenelse{\boolean{cms@external}}{
\begin{multline}
\label{azimuthal}
\frac{\rd N}{\rd\phi} \propto  1 + 2\sum_{n} \bigl(v_{n}\cos[n(\phi-\Psi_\mathrm{RP})] \\
+ a_{n}\sin[n(\phi-\Psi_\mathrm{RP})]\bigr),
\end{multline}
}{\begin{equation}
\label{azimuthal}
\frac{\rd N}{\rd\phi} \propto  1 + 2\sum_{n} \bigl(v_{n}\cos[n(\phi-\Psi_\mathrm{RP})] + a_{n}\sin[n(\phi-\Psi_\mathrm{RP})]\bigr),
\end{equation}
}
\end{linenomath}
where $\phi - \Psi_\mathrm{RP}$ represents the particle azimuthal angle with respect to the reaction plane angle $\Psi_\mathrm{RP}$ (determined by the impact parameter and beam axis), $v_{n}$ and $a_{n}$ denote the coefficients of $P$-even and $P$-odd Fourier terms, respectively. Although the reaction plane is not an experimental observable, it can be approximated by the second-order event
plane, $\Psi_\mathrm{EP}$, determined by the direction of the beam and the maximal particle
density in the elliptic azimuthal anisotropy. An azimuthal correlator
proposed to explore the first coefficient, $a_1$, of the $P$-odd Fourier terms
characterizing the charge separation~\cite{Voloshin:2004vk} is:
\begin{linenomath}
\ifthenelse{\boolean{cms@external}}{
\begin{multline}
\label{2pcorrelatorEP}
\left\langle \cos(\phi_{\alpha} + \phi_{\beta} - 2\Psi_\mathrm{EP}) \right\rangle  = \\
\langle \cos(\phi_{\alpha}-\Psi_\mathrm{EP})
\cos(\phi_{\beta}-\Psi_\mathrm{EP}) \rangle\\
- \langle \sin(\phi_{\alpha}-\Psi_\mathrm{EP})\sin(\phi_{\beta}-\Psi_\mathrm{EP}) \rangle.
\end{multline}
}{
\begin{multline}
\label{2pcorrelatorEP}
\left\langle \cos(\phi_{\alpha} + \phi_{\beta} - 2\Psi_\mathrm{EP}) \right\rangle  = \left\langle \cos(\phi_{\alpha}-\Psi_\mathrm{EP})\cos(\phi_{\beta}-\Psi_\mathrm{EP}) \right\rangle \\
- \left\langle \sin(\phi_{\alpha}-\Psi_\mathrm{EP})\sin(\phi_{\beta}-\Psi_\mathrm{EP}) \right\rangle.
\end{multline}
}
\end{linenomath}
Here, $\alpha$ and $\beta$ denote particles with the same or opposite charge sign and the brackets reflect an averaging over particles and events. Assuming particles $\alpha, \beta$ are uncorrelated except for their individual correlations
with respect to the event plane, the first term on the right hand side of Eq.~(\ref{2pcorrelatorEP})
becomes $\left\langle v_{1,\alpha}v_{1,\beta} \right\rangle$, which is generally small
and independent of charge~\cite{Abelev:2009ad}, while the second term is sensitive to
charge separation and can be expressed as $\left\langle a_{1,\alpha}a_{1,\beta} \right\rangle$,
which can be measured.

The observation of the CME in heavy ion collisions remains
inconclusive because of several identified sources of background
correlations that can account for part or all of the observed charge-dependent
azimuthal correlations~\cite{Wang:2009kd,Bzdak:2010fd,Schlichting:2010qia}.
For example, the effect of local charge conservation, coupled with the anisotropic emission of particles ($v_{2}$), can generate an effect resembling charge separation with respect to the reaction plane~\cite{Schlichting:2010qia}.
The charge-dependent azimuthal correlation signals observed in the data can be qualitatively described by models that do not include CME, such as the AMPT~\cite{Ma:2011uma} and \textsc{epos lhc}~\cite{Pierog:2013ria} models.
A significant amount of recent experimental
and theoretical effort is directed toward quantifying possible mechanisms, including the CME, that
can lead to charge-dependent azimuthal correlations~\cite{Kharzeev:2015znc}.

This Letter presents the first application of charge-dependent azimuthal correlation analysis with respect to the event plane in proton-nucleus collisions, using \pPb data collected with the CMS detector at the LHC at $\rootsNN = 5.02\TeV$.
High-multiplicity \pp and \pPb collisions have been shown to generate large final-state azimuthal anisotropies, comparable to those in \AonA  collisions~\cite{Khachatryan:2010gv,Aad:2015gqa,Khachatryan:2015lva,Khachatryan:2016txc,CMS:2012qk,alice:2012qe,atlas:2012fa,Aaij:2015qcq,Khachatryan:2014jra,ABELEV:2013wsa,Chatrchyan:2013nka,Aad:2014lta,Khachatryan:2015waa,Dusling:2015gta}.
However, the CME contribution to any charge-dependent signal is expected to be small
in a high-multiplicity \pPb collision, as the proton likely intersects the Pb nucleus
at a small impact parameter. Consequently, the magnetic field in the proton-nucleus overlap
region is expected to be smaller than in peripheral \PbPb  collisions at similar multiplicities.
Furthermore, based on Monte Carlo (MC) Glauber calculations~\cite{Alver:2008aq},
the angle between the magnetic field direction and the event plane of elliptic anisotropy is randomly distributed in \pPb collisions, contrary to the situation for \PbPb  collisions.
With a reduced magnetic field strength and a random field orientation, the CME contribution to any
charge-dependent signal is expected to be small. The high-multiplicity events in \pPb collisions
exhibit collective effects and bulk properties similar to those found in \AonA  collisions~\cite{Chatrchyan:2013nka, Khachatryan:2016yru, Khachatryan:2015waa} but possess very different strengths and configurations of the initial magnetic field. Thus, they can provide a new way to explore the possible CME and local strong parity violation.
With the implementation of a high-multiplicity trigger,
the \pPb data sample gives access to multiplicities comparable to those in peripheral
\PbPb\ collisions (\eg, $\sim$55\% centrality, where centrality is defined as the fraction of the total
inelastic cross section, with 0\% denoting the most central collisions), allowing for a direct comparison of the two systems with
very different CME contributions in the overlap zone.
The measurement is presented in different charge combinations as functions of
event multiplicity and pseudorapidity ($\eta$) difference of correlated particles. In \pPb collisions,
the particle correlations with respect to the event planes that are obtained
using particles with $4.4<\abs{\eta}<5$ from the \Pp- and Pb-going beam direction, are also explored.

The central feature of the CMS apparatus is a superconducting solenoid of 6\unit{m}
internal diameter, providing a magnetic field of 3.8\unit{T}. Within the solenoid volume,
there are four primary subdetectors including a silicon
pixel and strip tracker detector, a lead tungstate crystal electromagnetic calorimeter
(ECAL), and a brass and scintillator hadron calorimeter (HCAL), each composed of a
barrel and two endcap sections. Muons are measured in gas-ionization detectors
embedded in the steel flux-return yoke outside the solenoid. The silicon tracker
measures charged particles within the range $\abs{\eta}< 2.5$. For charged
particles with transverse momentum $1 < \pt < 10\GeV$ and $\abs{\eta} < 1.4$, the track resolutions are
typically 1.5\% in \pt and 25--90 (45--150)\mum in the transverse (longitudinal)
impact parameter~\cite{TRK-11-001}. Iron and quartz-fiber Cherenkov hadron forward
(HF) calorimeters cover the range $2.9 < \abs{\eta} < 5.2$. A detailed description of the CMS detector,
together with a definition of the coordinate system used and the relevant kinematic variables, can be found in
Ref.~\cite{Chatrchyan:2008zzk}.

The \pPb data at $\rootsNN = 5.02\TeV$, collected in 2013 at the LHC, correspond to an integrated luminosity of $35\unit{nb}^{-1}$. The beam energies are 4\TeV for the protons and 1.58\TeV per nucleon for the lead nuclei.
 A subset of peripheral \PbPb  data at $\rootsNN = 5.02\TeV$ collected
in 2015 (30--80\% centrality) is also used.
The \PbPb  data were reprocessed using the same reconstruction algorithm
as the \pPb data, in order to directly compare the two systems at similar multiplicities.
The event reconstruction, event selections, and the triggers, including the dedicated triggers
to collect a large sample of high-multiplicity \pPb events, are identical to those used in previous
CMS particle correlation measurements~\cite{Chatrchyan:2013nka,Khachatryan:2010gv}.
 In the offline analysis of \pPb (\PbPb) collisions, hadronic events
are selected by requiring the presence of at least one (three) energy deposit(s) greater
than 3\GeV in each of the two HF calorimeters. Events are also required to
contain a primary vertex within 15\unit{cm} of the nominal interaction point
along the beam axis and 0.15\unit{cm} in the transverse direction.
 In the \pPb data sample,
there is a 3\% probability to have at least one additional interaction in
the same bunch crossing (pileup). The procedure used to reject pileup events
in \pPb collisions yields a purity of 99.8\% for single pPb collision events and is described in Ref.~\cite{Chatrchyan:2013nka}. The pileup in \PbPb  data
is negligible.

Primary tracks, i.e., tracks that originate at the primary
vertex and satisfy the high-purity criteria of Ref.~\cite{TRK-11-001},
are used to define the event charged-particle multiplicity (\noff) and to perform correlation
measurements. In addition, the impact parameter significance of the track with
respect to the primary vertex in the direction along the beam axis, $d_z/\sigma(d_z)$
is required to be less than 3, as is the corresponding impact parameter significance in the
transverse plane, $d_\mathrm{T}/\sigma(d_\mathrm{T})$. The relative uncertainty in \pt,
$\sigma(\pt)/\pt$, must be less than 10\%. Each track is also required to leave
at least one hit in one of the three layers of the pixel tracker.
To ensure high tracking efficiency, only tracks with
$\abs{\eta}<2.4$ and $\pt > 0.3\GeV$ are used in this analysis.

The \pPb and \PbPb  data are compared in classes of \noff,
where primary tracks with $\abs{\eta}<2.4$ and $\pt >0.4$\GeV are counted. To compare
with results from other experiments, the \PbPb  data are also analyzed based on centrality
classes for the 30--80\% centrality range.
The average values of multiplicity, before and after correcting for
detector and algorithm inefficiencies, in each multiplicity class
of \pPb and \PbPb  data, can be found in Ref.~\cite{Chatrchyan:2013nka}.

Without directly reconstructing the event plane, the expression shown in
Eq.~(\ref{2pcorrelatorEP}) can be alternatively evaluated using a three-particle
correlator with respect to a third particle~\cite{Abelev:2009ac,Abelev:2009ad},
$\left\langle \cos(\phi_{\alpha} + \phi_{\beta} - 2\phi_{c}) \right\rangle / v_{2,c}$,
where $v_{2,c}$ corresponds to the elliptic flow of the particle $c$.
The three-particle correlator is measured via the scalar product method
of $Q$ vectors~\cite{Abelev:2009ad, Bilandzic:2013kga}.
The particles $\alpha$ and $\beta$ are taken from the tracker with
$\abs{\eta} < 2.4$ and $0.3<\pt<3$\GeV, and are corrected for tracking efficiency to account for reconstruction effects.
 The particle $c$ is measured by using the tower energies in the HF calorimeters
with $4.4 < \abs{\eta} < 5.0$. This choice of $\eta$ range for HF towers
imposes an $\eta$ gap of at least 2 units with respect to particles $\alpha$
and $\beta$ from the tracker, to minimize possible short-range correlations.
To account for any occupancy effect of the HF detectors resulting from the large
granularities in $\eta$ and $\phi$, each tower is weighted by its \ET value when calculating the $Q$ vector.
The $v_{2,c}$ is obtained following the standard scalar-product method~\cite{Abelev:2009ac,Abelev:2009ad}, by correlating the Q vectors from the tracker region at midrapidity and the two HF detectors at forward rapidity.
 The three-particle correlator is evaluated for particles $\alpha$ and $\beta$
carrying same sign (SS) and opposite sign (OS), as a function of pseudorapidity
difference \deta\ ($\equiv \abs{\eta_{\alpha}-\eta_{\beta}}$). The SS combinations,
($+,+$) and ($-,-$), give consistent results within statistical uncertainty and are therefore combined. For \pPb collisions, the three-particle correlator is also measured with particle $c$
from HF$+$ and HF$-$, corresponding to the \Pp- and Pb-going direction, respectively.
For symmetric \PbPb  collisions, the results from HF+ and HF$-$ are consistent with each other within statistical uncertainty and are therefore averaged. The effect of the nonuniform detector acceptance is found to be negligible by evaluating
the cumulants of $Q$-vector products~\cite{Selyuzhenkov:2007zi}.

The absolute systematic uncertainty of the three-particle correlator has been studied. Varying
the $d_z/\sigma(d_z)$ and $d_\mathrm{T}/\sigma(d_\mathrm{T})$ from less than 3
(default) to less than 2 and 5, and the $\sigma(\pt)/\pt < 10\%$ (default) to
$\sigma(\pt)/\pt < 5\%$, together yield a systematic uncertainty of ${\pm}1.0\times 10^{-5}$.
The longitudinal primary vertex position ($V_z$) has been varied, using ranges
$\abs{V_{z}} < 3$\unit{cm} and $3 < \abs{V_z} < 15$\unit{cm}, where the difference with respect to the default range $\abs{V_z} < 15$\unit{cm} is ${\pm}1.0\times 10^{-5}$, taken as the systematic uncertainty.
In \pPb collisions only, using the lower-threshold of the high-multiplicity trigger yields a systematic uncertainty of $\pm 3.0\times 10^{-5}$, which accounts for the possible trigger bias from the inefficiency of the default trigger around the threshold. A final test of the analysis procedures is
done by comparing ``known" charge-dependent signals based on the \textsc{epos}
event generator to those found after events are passed through a \GEANTfour~\cite{GEANT4}
simulation of the CMS detector response. Based on this test, a systematic uncertainty of
${\pm}2.5\times 10^{-5}$ is assigned. The tracking efficiency and acceptance of positively and negatively charged particles have been evaluated separately, and the difference has been found to be negligible.
All sources of systematic uncertainty are uncorrelated and added in quadrature to obtain
the total absolute systematic uncertainty. No dependence of the systematic uncertainties on the
sign combination, multiplicity, or $\Delta\eta$ is found. The systematic uncertainties in our results as function of \deta and multiplicity are point-to-point correlated. In \pPb collisions, the systematic uncertainty is also observed to be independent of particle $c$ pointing to the Pb- or \Pp-going direction, and
thus is quoted to be the same for these two situations.

\begin{figure*}[thb]
\centering
\includegraphics[width=0.85\textwidth]{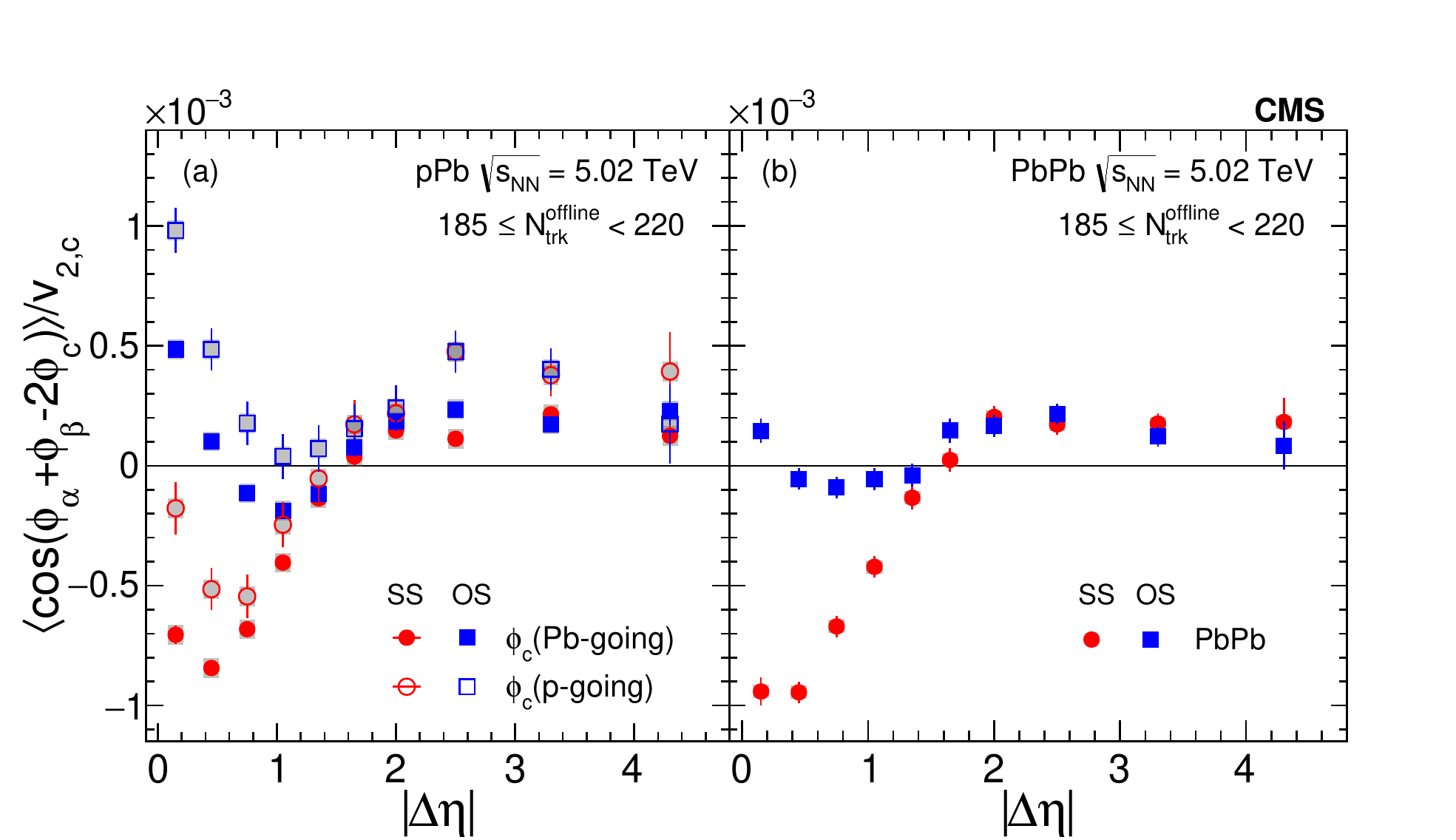}
\caption{
The same (SS) and opposite sign (OS) three-particle correlator
as a function of $\deta \equiv \abs{\eta_{\alpha}-\eta_{\beta}}$ for $185 \leq \noff < 220$
in (a)~\pPb and (b)~\PbPb  collisions at $\rootsNN = 5.02\TeV$.
The \pPb results obtained with particle $c$ in Pb-going (solid markers)
and \Pp-going (open markers) sides are shown separately.
Statistical and systematic uncertainties
are indicated by the error bars and shaded regions, respectively.}
\label{fig:figure1}
\end{figure*}

Measurements of the charge-dependent three-particle correlator
are shown in Fig.~\ref{fig:figure1} as a function of the \deta between charged particles $\alpha$ and $\beta$
with the same and opposite signs, in the multiplicity range $185 \leq \noff < 220$
for \pPb and \PbPb  collisions at $\rootsNN = 5.02\TeV$. The \pPb data are
obtained with particle $c$ in the Pb- and \Pp-going
sides separately. In both \pPb and \PbPb  systems,
a charge dependence of the three-particle correlator is observed for $\deta$
up to about 1.6. In this range, the SS correlators show significant negative values as $\deta$ decreases,
while the OS correlators become positive towards $\deta \approx 0$. For $\deta > 1.6$,
the SS and OS correlators converge to a common positive value,
which is weakly dependent on $\deta$ up to about 4.8 units. Similar $\deta$
dependence of the three-particle correlator has been reported at $\rootsNN =
0.2\TeV$~\cite{Abelev:2009ac} and 2.76\TeV~\cite{Abelev:2012pa}, measured up to $\deta \approx 1.6$.
In \pPb collisions, three-particle correlators obtained with particle $c$ from the \Pp-going side
are shifted toward more positive values than those from the Pb-going side by approximately the same amount for both the SS and OS pairs. The Pb-going side results for the \pPb collisions are of
similar magnitude as the results for \PbPb  collisions. The common shift of SS and OS correlators between the \Pp- and Pb-going side reference ($c$) particle, may be related to
sources of correlations that are charge-independent, such as directed flow and the momentum
conservation effect, the latter being sensitive to the difference in multiplicity between
\Pp- and Pb-going directions.

\begin{figure}[thb]
\centering
\includegraphics[width=0.46\textwidth]{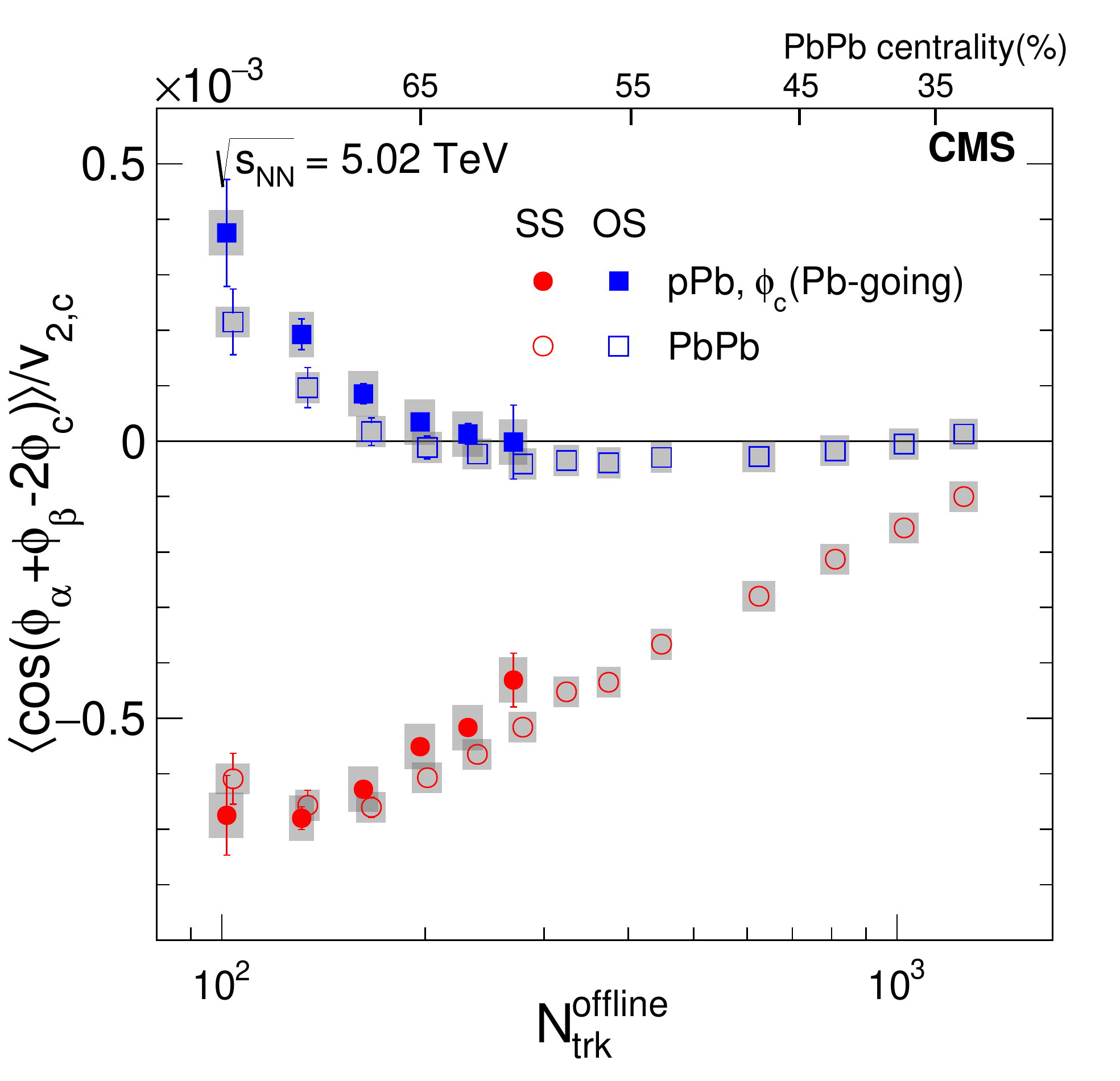}
 \caption{
The same sign (SS) and opposite sign (OS) three-particle correlator averaged over
$\abs{\eta_{\alpha}-\eta_{\beta}}<1.6$ as a function of \noff in \pPb and \PbPb  collisions at $\rootsNN = 5.02\TeV$ are
shown. Statistical and systematic uncertainties are indicated by the error bars and shaded regions, respectively.}
\label{fig:figure2}
\end{figure}

To explore the multiplicity or centrality dependence
of the three-particle correlator, an average of the results in Fig.~\ref{fig:figure1}
over $\deta<1.6$ (charge-dependent region) is taken, where the average is weighted by the number of particle pairs in each $\deta$ range. The resulting $\deta$-averaged three-particle
correlators are shown in Fig.~\ref{fig:figure2} as a function of \noff for \pPb (particle $c$ from the Pb-going side) and \PbPb  collisions at $\rootsNN = 5.02$\TeV.
Up to $\noff = 300$, the \pPb and \PbPb  results are measured in the same \noff ranges. The centrality scale on the top of
Fig.~\ref{fig:figure2} relates to
the \PbPb  experimental results.
Within uncertainties, the SS and OS correlators in \pPb and \PbPb  collisions
exhibit the same magnitude and trend as a function of event multiplicity. The OS
correlator reaches a value close to zero for $\noff > 200$, while the
SS correlator remains negative, but the magnitude gradually decreases as
\noff increases. Part of the observed multiplicity (or centrality) dependence is understood as a dilution effect that falls with the inverse of event multiplicity~\cite{Abelev:2009ad}.
The notably similar magnitude and multiplicity dependence
of the three-particle correlator observed in \pPb collisions relative to that in \PbPb
collisions again indicates that the dominant contribution
of the signal is not related to the CME.
The results of SS and OS three-particle correlators as functions of centrality in \PbPb collisions at $\rootsNN = 5.02$\TeV are also found to be consistent with the results from
lower energy \AonA\ collisions~\cite{Abelev:2009ad, Abelev:2012pa}.

\begin{figure}[thb]
\centering
\includegraphics[width=0.48\textwidth]{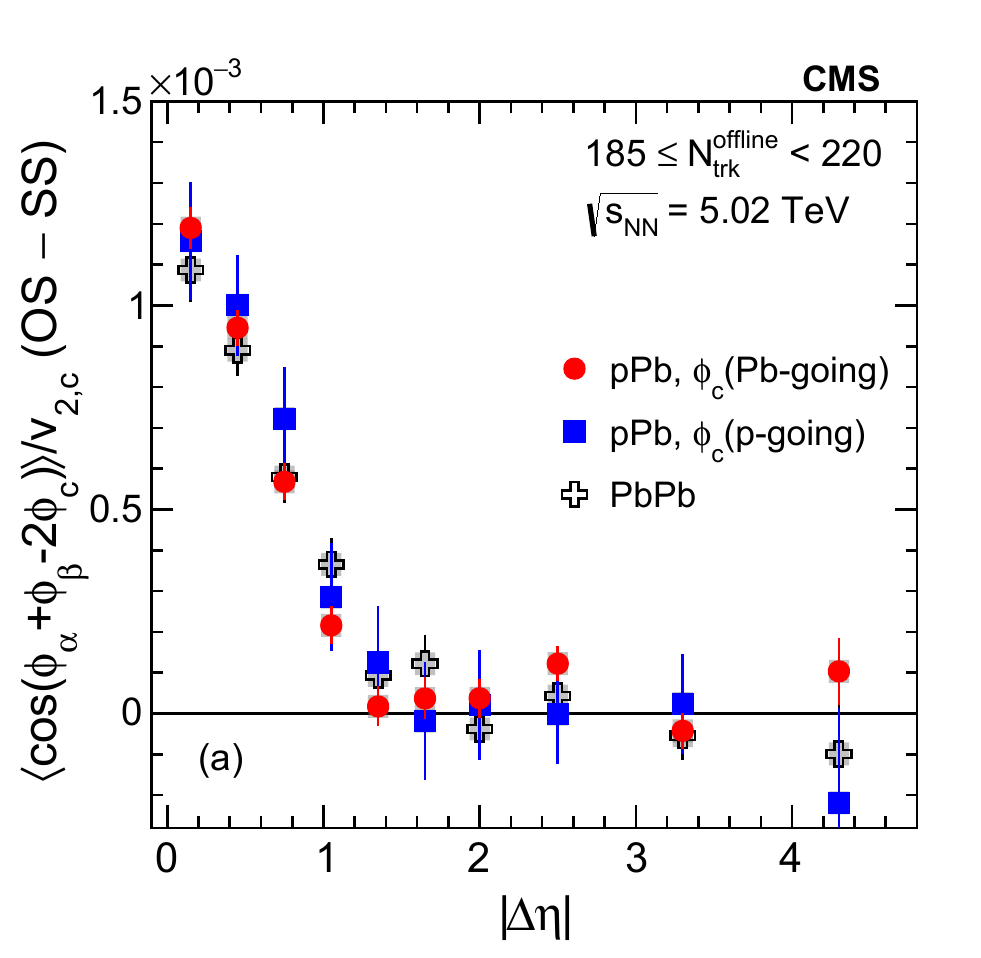}
\includegraphics[width=0.48\textwidth]{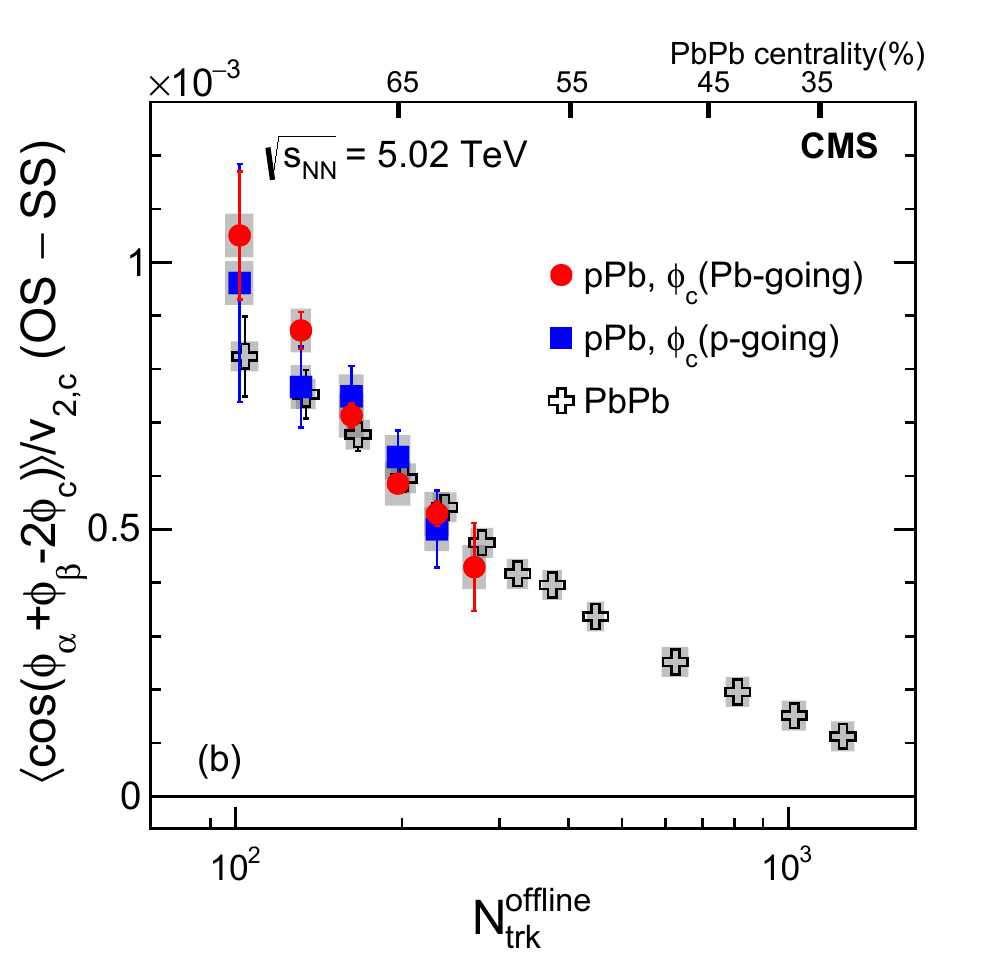}
 \caption{
The difference of the opposite sign (OS) and same sign (SS) three-particle correlators
(a)~as a function of $\abs{\eta_{\alpha}-\eta_{\beta}}$ for
$185 \leq \noff < 220$ and (b)~as a function of \noff, averaged over
$\abs{\eta_{\alpha}-\eta_{\beta}}<1.6$, in \pPb and \PbPb  collisions at $\rootsNN = 5.02$\TeV.
The \pPb results are obtained with particle $c$ from Pb- and \Pp-going
sides separately. Statistical and systematic uncertainties
are indicated by the error bars and shaded regions, respectively.
}
\label{fig:figure3}
\end{figure}

To eliminate sources of correlations that are charge independent
(\eg, directed flow, $v_1$) and to explore a possible charge separation effect
generated by the CME, the difference of three-particle correlators between
OS and SS is shown as a function of $\deta$
in the multiplicity range $185 \leq \noff < 220$ (Fig.~\ref{fig:figure3} (a))
and as a function of \noff averaged over $\deta< 1.6$ (Fig.~\ref{fig:figure3} (b),)
for \pPb and \PbPb  collisions at $\rootsNN = 5.02\TeV$.
After taking the difference, the \pPb data
with particle $c$ from both the
p- and Pb-going sides, and \PbPb  data, show nearly identical
values.
The charge-dependent difference is largest at $\deta \approx 0$ and drops to zero for $\deta > 1.6$, and also decreases as a function of \noff. The striking similarity in the observed charge-dependent
azimuthal correlations strongly suggests a common physical origin. In \PbPb
collisions, it was suggested that the charge dependence of the three-particle
correlator as well as its $\deta$ dependence are indications of the charge separation
effect with respect to the event plane due to the CME~\cite{Abelev:2009ad, Abelev:2012pa}.
However, as argued earlier, a strong charge separation signal from the CME
is not expected in a very high-multiplicity \pPb collision.
The similarity
seen between high-multiplicity \pPb and peripheral \PbPb collisions
challenges the attribution of the observed charge-dependent correlations to
the CME. Note that there is a hint of a slight difference between \pPb and \PbPb in the slopes of the \noff dependence in Fig.~\ref{fig:figure3} (b), where the systematic uncertainties are point-to-point correlated. This difference is worth further investigation.

In summary, charge-dependent azimuthal correlations of same and opposite sign
particles with respect to the second-order event plane
have been measured in \pPb and \PbPb
collisions at $\rootsNN = 5.02\TeV$ by the CMS experiment at the LHC.
The correlation is extracted via a three-particle correlator
as functions of particle \deta
and charged-particle multiplicity of the event.
The difference between opposite and same sign particles as functions of \deta  and multiplicity is found
to agree for \pPb and \PbPb  collisions, possibly indicating a
common underlying mechanism that generates the observed correlation.
These results challenge the CME interpretation
for the observed charge-dependent azimuthal correlations in nucleus-nucleus collisions
at RHIC and the LHC.

\begin{acknowledgments}
We congratulate our colleagues in the CERN accelerator departments for the excellent performance of the LHC and thank the technical and administrative staffs at CERN and at other CMS institutes for their contributions to the success of the CMS effort. In addition, we gratefully acknowledge the computing centers and personnel of the Worldwide LHC Computing Grid for delivering so effectively the computing infrastructure essential to our analyses. Finally, we acknowledge the enduring support for the construction and operation of the LHC and the CMS detector provided by the following funding agencies: BMWFW and FWF (Austria); FNRS and FWO (Belgium); CNPq, CAPES, FAPERJ, and FAPESP (Brazil); MES (Bulgaria); CERN; CAS, MoST, and NSFC (China); COLCIENCIAS (Colombia); MSES and CSF (Croatia); RPF (Cyprus); SENESCYT (Ecuador); MoER, ERC IUT and ERDF (Estonia); Academy of Finland, MEC, and HIP (Finland); CEA and CNRS/IN2P3 (France); BMBF, DFG, and HGF (Germany); GSRT (Greece); OTKA and NIH (Hungary); DAE and DST (India); IPM (Iran); SFI (Ireland); INFN (Italy); MSIP and NRF (Republic of Korea); LAS (Lithuania); MOE and UM (Malaysia); BUAP, CINVESTAV, CONACYT, LNS, SEP, and UASLP-FAI (Mexico); MBIE (New Zealand); PAEC (Pakistan); MSHE and NSC (Poland); FCT (Portugal); JINR (Dubna); MON, RosAtom, RAS and RFBR (Russia); MESTD (Serbia); SEIDI and CPAN (Spain); Swiss Funding Agencies (Switzerland); MST (Taipei); ThEPCenter, IPST, STAR and NSTDA (Thailand); TUBITAK and TAEK (Turkey); NASU and SFFR (Ukraine); STFC (United Kingdom); DOE and NSF (USA).
\end{acknowledgments}
\bibliography{auto_generated}

\providecommand{\href}[2]{#2}\begingroup\raggedright\begin{thebibliography}{10}%
\makeatletter
\providecommand{\hrefCMSnoop }[0]{\@secondoftwo}%
\makeatother
\providecommand{\doi}{\texttt{doi:}\begingroup \urlstyle{tt}\Url}

\bibitem{Lee:1973iz}
\hrefCMSnoop {}{T.~D. Lee, ``A Theory of Spontaneous {$T$} Violation'',}
  \textit{ Phys. Rev. D} \textbf{ 8} (1973) 1226,
\href{http://dx.doi.org/10.1103/PhysRevD.8.1226}{\doi{10.1103/PhysRevD.8.1226}}.

\bibitem{Lee:1974ma}
\hrefCMSnoop {}{T.~D. Lee and G.~C. Wick, ``{Vacuum stability and vacuum
  excitation in a spin-0 field theory}'',} \textit{ Phys. Rev. D} \textbf{ 9}
  (1974) 2291,
\href{http://dx.doi.org/10.1103/PhysRevD.9.2291}{\doi{10.1103/PhysRevD.9.2291}}.

\bibitem{Morley1985}
\hrefCMSnoop {}{P.~D. Morley and I.~A. Schmidt, ``Strong P, CP, T violations in
  heavy-ion collisions'',} \textit{ Z. Phys. C} \textbf{ 26} (1985) 627,
  \href{http://dx.doi.org/10.1007/BF01551807}{\doi{10.1007/BF01551807}}.

\bibitem{Kharzeev:1998kz}
\hrefCMSnoop {}{D.~Kharzeev, R.~D. Pisarski, and M.~H.~G. Tytgat, ``Possibility
  of Spontaneous Parity Violation in Hot {QCD}'',} \textit{ Phys. Rev. Lett.}
  \textbf{ 81} (1998) 512,
  \href{http://dx.doi.org/10.1103/PhysRevLett.81.512}{\doi{10.1103/PhysRevLett.81.512}},
\href{http://www.arXiv.org/abs/hep-ph/9804221}{\texttt{arXiv:hep-ph/9804221}}.

\bibitem{Kharzeev:2004ey}
\hrefCMSnoop {}{D.~Kharzeev, ``Parity violation in hot {QCD}: Why it can
  happen, and how to look for it'',} \textit{ Phys. Lett. B} \textbf{ 633}
  (2006) 260,
  \href{http://dx.doi.org/10.1016/j.physletb.2005.11.075}{\doi{10.1016/j.physletb.2005.11.075}},
\href{http://www.arXiv.org/abs/hep-ph/0406125}{\texttt{arXiv:hep-ph/0406125}}.

\bibitem{Abelev:2009ac}
\hrefCMSnoop {}{{STAR} Collaboration, ``{Azimuthal Charged-Particle
  Correlations and Possible Local Strong Parity Violation}'',} \textit{ Phys.
  Rev. Lett.} \textbf{ 103} (2009) 251601,
  \href{http://dx.doi.org/10.1103/PhysRevLett.103.251601}{\doi{10.1103/PhysRevLett.103.251601}},
\href{http://www.arXiv.org/abs/0909.1739}{\texttt{arXiv:0909.1739}}.

\bibitem{Abelev:2009ad}
\hrefCMSnoop {}{{STAR} Collaboration, ``{Observation of charge-dependent
  azimuthal correlations and possible local strong parity violation in
  heavy-ion collisions}'',} \textit{ Phys. Rev. C} \textbf{ 81} (2010) 054908,
  \href{http://dx.doi.org/10.1103/PhysRevC.81.054908}{\doi{10.1103/PhysRevC.81.054908}},
\href{http://www.arXiv.org/abs/0909.1717}{\texttt{arXiv:0909.1717}}.

\bibitem{Adamczyk:2013kcb}
\hrefCMSnoop {}{{STAR} Collaboration, ``{Measurement of charge multiplicity
  asymmetry correlations in high-energy nucleus-nucleus collisions at \rootsNN\
  = 200\GeV}'',} \textit{ Phys. Rev. C} \textbf{ 89} (2014) 044908,
  \href{http://dx.doi.org/10.1103/PhysRevC.89.044908}{\doi{10.1103/PhysRevC.89.044908}},
\href{http://www.arXiv.org/abs/1303.0901}{\texttt{arXiv:1303.0901}}.

\bibitem{Adamczyk:2014mzf}
\hrefCMSnoop {}{{STAR} Collaboration, ``{Beam-Energy Dependence of Charge
  Separation along the Magnetic Field in Au+Au Collisions at RHIC}'',} \textit{
  Phys. Rev. Lett.} \textbf{ 113} (2014) 052302,
  \href{http://dx.doi.org/10.1103/PhysRevLett.113.052302}{\doi{10.1103/PhysRevLett.113.052302}},
\href{http://www.arXiv.org/abs/1404.1433}{\texttt{arXiv:1404.1433}}.

\bibitem{Adamczyk:2013hsi}
\hrefCMSnoop {}{{STAR} Collaboration, ``{Fluctuations of charge separation
  perpendicular to the event plane and local parity violation in \rootsNN\ =
  200\GeV Au+Au collisions at the BNL Relativistic Heavy Ion Collider}'',}
  \textit{ Phys. Rev. C} \textbf{ 88} (2013) 064911,
  \href{http://dx.doi.org/10.1103/PhysRevC.88.064911}{\doi{10.1103/PhysRevC.88.064911}},
\href{http://www.arXiv.org/abs/1302.3802}{\texttt{arXiv:1302.3802}}.

\bibitem{Abelev:2012pa}
\hrefCMSnoop {}{{ALICE} Collaboration, ``{Charge separation relative to the
  reaction plane in Pb-Pb collisions at \rootsNN\ = 2.76\TeV}'',} \textit{
  Phys. Rev. Lett.} \textbf{ 110} (2013) 012301,
  \href{http://dx.doi.org/10.1103/PhysRevLett.110.012301}{\doi{10.1103/PhysRevLett.110.012301}},
\href{http://www.arXiv.org/abs/1207.0900}{\texttt{arXiv:1207.0900}}.

\bibitem{Voloshin:2004vk}
\hrefCMSnoop {}{S.~A. Voloshin, ``{Parity violation in hot QCD: How to detect
  it}'',} \textit{ Phys. Rev. C} \textbf{ 70} (2004) 057901,
  \href{http://dx.doi.org/10.1103/PhysRevC.70.057901}{\doi{10.1103/PhysRevC.70.057901}},
\href{http://www.arXiv.org/abs/hep-ph/0406311}{\texttt{arXiv:hep-ph/0406311}}.

\bibitem{Wang:2009kd}
\hrefCMSnoop {}{F.~Wang, ``{Effects of cluster particle correlations on local
  parity violation observables}'',} \textit{ Phys. Rev. C} \textbf{ 81} (2010)
  064902,
  \href{http://dx.doi.org/10.1103/PhysRevC.81.064902}{\doi{10.1103/PhysRevC.81.064902}},
\href{http://www.arXiv.org/abs/0911.1482}{\texttt{arXiv:0911.1482}}.

\bibitem{Bzdak:2010fd}
\hrefCMSnoop {}{A.~Bzdak, V.~Koch, and J.~Liao, ``{Azimuthal correlations from
  transverse momentum conservation and possible local parity violation}'',}
  \textit{ Phys. Rev. C} \textbf{ 83} (2011) 014905,
  \href{http://dx.doi.org/10.1103/PhysRevC.83.014905}{\doi{10.1103/PhysRevC.83.014905}},
\href{http://www.arXiv.org/abs/1008.4919}{\texttt{arXiv:1008.4919}}.

\bibitem{Schlichting:2010qia}
\hrefCMSnoop {}{S.~Schlichting and S.~Pratt, ``{Charge conservation at energies
  available at the BNL Relativistic Heavy Ion Collider and contributions to
  local parity violation observables}'',} \textit{ Phys. Rev. C} \textbf{ 83}
  (2011) 014913,
  \href{http://dx.doi.org/10.1103/PhysRevC.83.014913}{\doi{10.1103/PhysRevC.83.014913}},
\href{http://www.arXiv.org/abs/1009.4283}{\texttt{arXiv:1009.4283}}.

\bibitem{Ma:2011uma}
\hrefCMSnoop {}{G.-L. Ma and B.~Zhang, ``{Effects of final state interactions
  on charge separation in relativistic heavy ion collisions}'',} \textit{ Phys.
  Lett. B} \textbf{ 700} (2011) 39,
  \href{http://dx.doi.org/10.1016/j.physletb.2011.04.057}{\doi{10.1016/j.physletb.2011.04.057}},
\href{http://www.arXiv.org/abs/1101.1701}{\texttt{arXiv:1101.1701}}.

\bibitem{Pierog:2013ria}
T.~Pierog\hrefCMSnoop {}{ {et~al.}, ``{EPOS LHC: Test of collective
  hadronization with data measured at the CERN Large Hadron Collider}'',}
  \textit{ Phys. Rev. C} \textbf{ 92} (2015) 034906,
  \href{http://dx.doi.org/10.1103/PhysRevC.92.034906}{\doi{10.1103/PhysRevC.92.034906}}.

\bibitem{Kharzeev:2015znc}
\hrefCMSnoop {}{D.~E. Kharzeev, J.~Liao, S.~A. Voloshin, and G.~Wang, ``{Chiral
  magnetic and vortical effects in high-energy nuclear collisions---A status
  report}'',} \textit{ Prog. Part. Nucl. Phys.} \textbf{ 88} (2016) 1,
  \href{http://dx.doi.org/10.1016/j.ppnp.2016.01.001}{\doi{10.1016/j.ppnp.2016.01.001}},
\href{http://www.arXiv.org/abs/1511.04050}{\texttt{arXiv:1511.04050}}.

\bibitem{Khachatryan:2010gv}
\hrefCMSnoop {}{{CMS} Collaboration, ``{Observation of long-range, near-side
  angular correlations in proton-proton collisions at the LHC}'',} \textit{
  JHEP} \textbf{ 09} (2010) 091,
  \href{http://dx.doi.org/10.1007/JHEP09(2010)091}{\doi{10.1007/JHEP09(2010)091}},
\href{http://www.arXiv.org/abs/1009.4122}{\texttt{arXiv:1009.4122}}.

\bibitem{Aad:2015gqa}
\hrefCMSnoop {}{{ATLAS} Collaboration, ``Observation of Long-Range Elliptic
  Azimuthal Anisotropies in $\roots = 13$ and 2.76 {TeV} $pp$ Collisions with
  the {ATLAS} Detector'',} \textit{ Phys. Rev. Lett.} \textbf{ 116} (2016)
  172301,
  \href{http://dx.doi.org/10.1103/PhysRevLett.116.172301}{\doi{10.1103/PhysRevLett.116.172301}},
\href{http://www.arXiv.org/abs/1509.04776}{\texttt{arXiv:1509.04776}}.

\bibitem{Khachatryan:2015lva}
\hrefCMSnoop {}{{CMS} Collaboration, ``Measurement of Long-Range Near-Side
  Two-Particle Angular Correlations in $pp$ collisions at $\roots = 13$
  {TeV}'',} \textit{ Phys. Rev. Lett.} \textbf{ 116} (2016) 172302,
  \href{http://dx.doi.org/10.1103/PhysRevLett.116.172302}{\doi{10.1103/PhysRevLett.116.172302}},
\href{http://www.arXiv.org/abs/1510.03068}{\texttt{arXiv:1510.03068}}.

\bibitem{Khachatryan:2016txc}
\hrefCMSnoop {}{{CMS} Collaboration, ``Evidence for collectivity in pp
  collisions at the {LHC}'',} \textit{ Phys. Lett. B} \textbf{ 765} (2017) 193,
  \href{http://dx.doi.org/10.1016/j.physletb.2016.12.009}{\doi{10.1016/j.physletb.2016.12.009}},
\href{http://www.arXiv.org/abs/1606.06198}{\texttt{arXiv:1606.06198}}.

\bibitem{CMS:2012qk}
\hrefCMSnoop {}{{CMS} Collaboration, ``{Observation of long-range, near-side
  angular correlations in pPb collisions at the LHC}'',} \textit{ Phys. Lett.
  B} \textbf{ 718} (2013) 795,
  \href{http://dx.doi.org/10.1016/j.physletb.2012.11.025}{\doi{10.1016/j.physletb.2012.11.025}},
\href{http://www.arXiv.org/abs/1210.5482}{\texttt{arXiv:1210.5482}}.

\bibitem{alice:2012qe}
\hrefCMSnoop {}{{ALICE} Collaboration, ``Long-range angular correlations on the
  near and away side in {p-PB} collisions at {$\rootsNN = 5.02$\TeV}'',}
  \textit{ Phys. Lett. B} \textbf{ 719} (2013) 29,
  \href{http://dx.doi.org/10.1016/j.physletb.2013.01.012}{\doi{10.1016/j.physletb.2013.01.012}},
\href{http://www.arXiv.org/abs/1212.2001}{\texttt{arXiv:1212.2001}}.

\bibitem{atlas:2012fa}
\hrefCMSnoop {}{{ATLAS} Collaboration, ``Observation of Associated Near-Side
  and Away-Side Long-Range Correlations in {$\rootsNN = 5.02$\TeV} Proton-Lead
  Collisions with the {ATLAS} Detector'',} \textit{ Phys. Rev. Lett.} \textbf{
  110} (2013) 182302,
  \href{http://dx.doi.org/10.1103/PhysRevLett.110.182302}{\doi{10.1103/PhysRevLett.110.182302}},
\href{http://www.arXiv.org/abs/1212.5198}{\texttt{arXiv:1212.5198}}.

\bibitem{Aaij:2015qcq}
\hrefCMSnoop {}{{LHCb} Collaboration, ``Measurements of long-range near-side
  angular correlations in {$\rootsNN = 5\TeV$} proton-lead collisions in the
  forward region'',} \textit{ Phys. Lett. B} \textbf{ 762} (2016) 473,
  \href{http://dx.doi.org/10.1016/j.physletb.2016.09.064}{\doi{10.1016/j.physletb.2016.09.064}},
\href{http://www.arXiv.org/abs/1512.00439}{\texttt{arXiv:1512.00439}}.

\bibitem{Khachatryan:2014jra}
\hrefCMSnoop {}{{CMS} Collaboration, ``{Long-range two-particle correlations of
  strange hadrons with charged particles in pPb and PbPb collisions at LHC
  energies}'',} \textit{ Phys. Lett. B} \textbf{ 742} (2015) 200,
  \href{http://dx.doi.org/10.1016/j.physletb.2015.01.034}{\doi{10.1016/j.physletb.2015.01.034}},
\href{http://www.arXiv.org/abs/1409.3392}{\texttt{arXiv:1409.3392}}.

\bibitem{ABELEV:2013wsa}
\hrefCMSnoop {}{{ALICE} Collaboration, ``{Long-range angular correlations of
  pi, K and p in p--Pb collisions at \rootsNN\ = 5.02\TeV}'',} \textit{ Phys.
  Lett. B} \textbf{ 726} (2013) 164,
  \href{http://dx.doi.org/10.1016/j.physletb.2013.08.024}{\doi{10.1016/j.physletb.2013.08.024}},
\href{http://www.arXiv.org/abs/1307.3237}{\texttt{arXiv:1307.3237}}.

\bibitem{Chatrchyan:2013nka}
\hrefCMSnoop {}{{CMS} Collaboration, ``{Multiplicity and transverse momentum
  dependence of two- and four-particle correlations in \pPb\ and \PbPb\
  collisions}'',} \textit{ Phys. Lett. B} \textbf{ 724} (2013) 213,
  \href{http://dx.doi.org/10.1016/j.physletb.2013.06.028}{\doi{10.1016/j.physletb.2013.06.028}},
\href{http://www.arXiv.org/abs/1305.0609}{\texttt{arXiv:1305.0609}}.

\bibitem{Aad:2014lta}
\hrefCMSnoop {}{{ATLAS} Collaboration, ``{Measurement of long-range
  pseudorapidity correlations and azimuthal harmonics in \rootsNN\ = 5.02\TeV
  proton-lead collisions with the ATLAS detector}'',} \textit{ Phys. Rev. C}
  \textbf{ 90} (2014) 044906,
  \href{http://dx.doi.org/10.1103/PhysRevC.90.044906}{\doi{10.1103/PhysRevC.90.044906}},
\href{http://www.arXiv.org/abs/1409.1792}{\texttt{arXiv:1409.1792}}.

\bibitem{Khachatryan:2015waa}
\hrefCMSnoop {}{{CMS} Collaboration, ``Evidence for Collective Multiparticle
  Correlations in {$p$-Pb} Collisions'',} \textit{ Phys. Rev. Lett.} \textbf{
  115} (2015) 012301,
  \href{http://dx.doi.org/10.1103/PhysRevLett.115.012301}{\doi{10.1103/PhysRevLett.115.012301}},
\href{http://www.arXiv.org/abs/1502.05382}{\texttt{arXiv:1502.05382}}.

\bibitem{Dusling:2015gta}
\hrefCMSnoop {}{K.~Dusling, W.~Li, and B.~Schenke, ``{Novel collective
  phenomena in high-energy proton-proton and proton-nucleus collisions}'',}
  \textit{ Int. J. Mod. Phys. E} \textbf{ 25} (2016) 1630002,
  \href{http://dx.doi.org/10.1142/S0218301316300022}{\doi{10.1142/S0218301316300022}},
\href{http://www.arXiv.org/abs/1509.07939}{\texttt{arXiv:1509.07939}}.

\bibitem{Alver:2008aq}
\hrefCMSnoop {}{B.~Alver, M.~Baker, C.~Loizides, and P.~Steinberg, ``{The
  PHOBOS Glauber Monte Carlo}'',} (2008).
\href{http://www.arXiv.org/abs/0805.4411}{\texttt{arXiv:0805.4411}}.

\bibitem{Khachatryan:2016yru}
\hrefCMSnoop {}{{CMS} Collaboration, ``Multiplicity and rapidity dependence of
  strange hadron production in pp, {pPb}, and {PbPb} collisions at the
  {LHC}'',} \textit{ Phys. Lett. B} \textbf{ 768} (2017) 103,
  \href{http://dx.doi.org/10.1016/j.physletb.2017.01.075}{\doi{10.1016/j.physletb.2017.01.075}},
\href{http://www.arXiv.org/abs/1605.06699}{\texttt{arXiv:1605.06699}}.

\bibitem{TRK-11-001}
\hrefCMSnoop {}{{CMS} Collaboration, ``{Description and performance of track
  and primary-vertex reconstruction with the CMS tracker}'',} \textit{ JINST}
  \textbf{ 9} (2014) P10009,
  \href{http://dx.doi.org/10.1088/1748-0221/9/10/P10009}{\doi{10.1088/1748-0221/9/10/P10009}},
\href{http://www.arXiv.org/abs/1405.6569}{\texttt{arXiv:1405.6569}}.

\bibitem{Chatrchyan:2008zzk}
\hrefCMSnoop {}{{CMS} Collaboration, ``The {CMS} experiment at the {CERN}
  {LHC}'',} \textit{ JINST} \textbf{ 3} (2008) S08004,
\href{http://dx.doi.org/10.1088/1748-0221/3/08/S08004}{\doi{10.1088/1748-0221/3/08/S08004}}.

\bibitem{Bilandzic:2013kga}
A.~Bilandzic\hrefCMSnoop {}{ {et~al.}, ``{Generic framework for anisotropic
  flow analyses with multiparticle azimuthal correlations}'',} \textit{ Phys.
  Rev. C} \textbf{ 89} (2014), no.~6, 064904,
  \href{http://dx.doi.org/10.1103/PhysRevC.89.064904}{\doi{10.1103/PhysRevC.89.064904}},
\href{http://www.arXiv.org/abs/1312.3572}{\texttt{arXiv:1312.3572}}.

\bibitem{Selyuzhenkov:2007zi}
\hrefCMSnoop {}{I.~Selyuzhenkov and S.~Voloshin, ``{Effects of nonuniform
  acceptance in anisotropic flow measurement}'',} \textit{ Phys. Rev. C}
  \textbf{ 77} (2008) 034904,
  \href{http://dx.doi.org/10.1103/PhysRevC.77.034904}{\doi{10.1103/PhysRevC.77.034904}},
\href{http://www.arXiv.org/abs/0707.4672}{\texttt{arXiv:0707.4672}}.

\bibitem{GEANT4}
\hrefCMSnoop {}{{GEANT4} Collaboration, ``{GEANT4}---a simulation toolkit'',}
  \textit{ Nucl. Instrum. Meth. A} \textbf{ 506} (2003) 250,
\href{http://dx.doi.org/10.1016/S0168-9002(03)01368-8}{\doi{10.1016/S0168-9002(03)01368-8}}.

\end{thebibliography}\endgroup
\cleardoublepage \appendix\section{The CMS Collaboration \label{app:collab}}\begin{sloppypar}\hyphenpenalty=5000\widowpenalty=500\clubpenalty=5000\textbf{Yerevan Physics Institute,  Yerevan,  Armenia}\\*[0pt]
V.~Khachatryan, A.M.~Sirunyan, A.~Tumasyan
\vskip\cmsinstskip
\textbf{Institut f\"{u}r Hochenergiephysik,  Wien,  Austria}\\*[0pt]
W.~Adam, E.~Asilar, T.~Bergauer, J.~Brandstetter, E.~Brondolin, M.~Dragicevic, J.~Er\"{o}, M.~Flechl, M.~Friedl, R.~Fr\"{u}hwirth\cmsAuthorMark{1}, V.M.~Ghete, C.~Hartl, N.~H\"{o}rmann, J.~Hrubec, M.~Jeitler\cmsAuthorMark{1}, A.~K\"{o}nig, I.~Kr\"{a}tschmer, D.~Liko, T.~Matsushita, I.~Mikulec, D.~Rabady, N.~Rad, B.~Rahbaran, H.~Rohringer, J.~Schieck\cmsAuthorMark{1}, J.~Strauss, W.~Waltenberger, C.-E.~Wulz\cmsAuthorMark{1}
% ...
\vskip\cmsinstskip
% ...
1:~~Also at Vienna University of Technology, Vienna, Austria\\
% ...

\end{sloppypar}
\end{document}